\documentclass[draftcls]{IEEEtran}

\usepackage{cite}

\ifCLASSINFOpdf
   \usepackage[pdftex]{graphicx}
\else
   \usepackage[dvips]{graphicx}
\fi
\usepackage[cmex10]{amsmath}
\usepackage{array}

\ifCLASSOPTIONcompsoc
  \usepackage[caption=false,font=normalsize,labelfont=sf,textfont=sf]{subfig}
\else
  \usepackage[caption=false,font=footnotesize]{subfig}
\fi
\usepackage{url}

\begin{document}
%
\title{Long term characterization of voltage references}
%
%
%

\author{Hubert~Halloin,
        Pierre~Prat,
        and~Julien~Brossard
\thanks{H. Halloin,~P. Prat~and~Julien~Brossard are with the APC, AstroParticule et Cosmologie, Universit\'e Paris Diderot, 
             CNRS/IN2P3, CEA/Irfu, Observatoire de Paris, Sorbonne Paris Cit\'e, 
             10, rue Alice Domon et L\'eonie Duquet, 
             75205 Paris Cedex 13, France 
              e-mail: hubert.halloin@apc.univ-paris7.fr.}
\thanks{Manuscript received ; revised .}}

%
%

\markboth{}%
{Halloin \MakeLowercase{\textit{et al.}}: Long term characterization of voltage references}
%



\maketitle

\begin{abstract}
We report here the characterization (temperature coefficients and noise level) of selected voltage references in the frequency range from $10^{-5}$ to 10 Hz. The goal of this work is to update previous studies, with a characterization at lower frequencies, and find voltage references that may be suitable for the space-based interferometry mission eLISA. The requirements of relative output stability of 1~ppm/$\sqrt{\text{Hz}}$ down to 0.1 mHz were not met by any of the tested devices, but 4 references approaches the objective~: the AD587UQ, the MAX6126AASA50, the LT1021-BCN8-5 and the LT6655BHM. While the first three were already identified as potential devices in previous studies, the later is a new promising candidate using a different technology (bandgap).

\end{abstract}


%
\IEEEpeerreviewmaketitle

\section{Introduction}
%
%
%
%
\IEEEPARstart{T}{he} evolved Laser Interferometer Space Antenna is a space-based project aiming at detecting gravitational waves in the frequency range 0.1 mHz to 10 Hz. eLISA consists of 3 spacecraft in a nearly-equilateral configuration, orbiting around the sun and forming a laser interferometer with arm length of $10^6$ km \cite{ELISAWP2013}. 
The eLISA mission has been submitted as an L-class mission to the European Space Agency and recently proposed for launch in 2034. The expected performance of the distance measurement with eLISA (a few pm/$\sqrt{\text{Hz}}$) requires ultra-stable electronic devices in general. 

Unfortunately, manufacturers usually do not characterize their electronics components around the mHz. In order to test discrete components as well as electronic boards, the APC (AstroParticle and Cosmology) laboratory is currently developing a dedicated test facility aiming at measuring their intrinsic performance at very low frequencies. 

As a first test case for this facility, this paper describes the stability characterization of selected voltage references. Long term stability of voltage references is of crucial importance to ensure the fidelity of phase measurements with eLISA : they are, e.g., used as reference for analog to digital converters (ADC), for photodiodes bias or capacitive sensing of the inertial mass position.  

Previous studies performed on the noise properties of voltage references usually focused on their use for metrological experiments in laboratories \cite{Valdes:1993p11400,Witt:1995p11398,Witt:1997p11391, Sebela:1998p11401, Helisto:2000p11408}, using voltage standards (such as Fluke 732B) that are not suitable for space-based operations. Additionally, most of these works measured the noise characteristics in frequency ranges above 0.1 Hz ('short-term' stability) or on months and years timescales (for international comparisons and fidelity of voltage references).  

In the present paper, we report tests on 8 voltage references (see table \ref{Tab:ManChar}), that were selected for their intrinsic good performances (low noise) and low temperature coefficient (as reported on the manufacturer's data sheet, i.e. down to 0.1 Hz). Bandgap, buried zener diodes and XFET technologies are represented.

In the context of the LISA mission similar studies have already been performed a few years ago \cite{Heinzel:2006p11548,Nickerson:2009p11547,Fleddermann:2009p9044}. Three of the voltage references tested in these papers (AD587UQ, MAX6126AASA50+ and LT1021BCN8-5) have also been studied in the present work. Two of these references are already space qualified (AD587 and LT1021). The LT1021 and MAX6350 have also been tested by the Swiss Federal Institute of Technology in Zurich (ETH), which is responsible for the redundant centralized voltage reference onboard eLISA \cite{Mance:2013}. The MAX6350 has also been radiation tested. The reader can refer to \cite{Rax:1997p11593} and \cite{Franco:2005p11592} for the effect of radiations on voltage references. In particular, \cite{Franco:2005p11592} suggests that some XFET models (such as the AD43x family) might be more radiation tolerant than Zener and bandgap devices.
It is however beyond the scope of this paper to discuss the effect of the space environment (mainly radiations) on the voltage reference characteristics.

To be compatible with the eLISA requirements, the amplitude spectral density of the voltage reference output should be better than 1~ppm/$\sqrt{\text{Hz}}$$\times \sqrt{1+\left(\frac{0.1\ \text{mHz}}{f}\right)^4}$.

\begin{table}[t]
\centering 
\renewcommand{\arraystretch}{1.0}
\caption{Models of voltage references tested in the present paper, together with the implemented technology} 
\begin{tabular}{c l l l}
\hline
 Number & Manufacturer & Model & Technology \\
\hline
1 & Analog Devices  & AD587UQ & buried zener \\
2 & Analog Devices  & ADR445BRZ & XFET \\
3 & Analog Devices  & ADR435BRZ & XFET\\
4 & Maxim Integrated & MAX6126AASA50 & proprietary \\
5 & Linear Technology & LTC6655BHMS8-5 & bandgap\\
6 & Maxim Integrated & MAX6350CSA+ & buried zener \\
7 & Linear Technology & LT1021BCN8-5 & buried zener\\
8 & Apex Microtechnology & VRE305AD & buried zener\\
\hline
\end{tabular}
\label{Tab:RefList}
\end{table}

\begin{table}[t]
\centering 
\renewcommand{\arraystretch}{1.0}
\caption{Electrical characteristics of voltage references (from the manufacturer data sheets). The nominal value of the output voltage is 5.0~V, except for the AD587 (10~V). All numerical values have been rescaled to ppm and the peak-to-peak noise level is given for the 0.1 to 10 Hz frequency range.} 
\begin{tabular}{c r r r r }
\hline
 &  Temp. drift & Line reg. & Noise & Long term\\
Num. & (typ-max)  &  (typ-max) & (typ) & drift\\
& [ppm/K] &  [ppm/V] & [ppm$_{p-p}$]  & [ppm/kHr]\\
\hline
1 & $<$5 & $<$10 & 0.4 & 15\\
2 & 1-3 & 10-20  & 0.45 & 50\\
3 & 1-3 & 5-20  & 1.6 & 40\\
4  & 0.5-3 & 0.6-8  & 0.6 & 20\\
5  & 1-2 & 5-25 & 0.25 & 60\\
6  & 0.5-1 & 2-5 & 0.6 & 30\\
7  & 2-5 & 2-6 & 0.6 & 15\\
8 & $<0.6$ & 6-10 & 0.6 & 6\\
\hline
\end{tabular}
\label{Tab:ManChar}
\end{table}

\section{Experimental setup}

The experimental setup and measurement procedures are driven by the main foreseeable noise sources in laboratory conditions, as well as the goal of the study~: the characterization of the intrinsic noise level between 0.1 mHz and 10 Hz and its comparison with eLISA requirements.
The main foreseen noise sources in the present experiment are the thermal drift of the output voltage, the influence of the power supply stability and the intrinsic noise of the measurement chain (amplifier and voltmeter). Pressure and hygrometry may also have an effect on the voltage stability \cite{Witt:1995p11398, Witt:1999p11395}.

\begin{figure}[!t]
\centering
\includegraphics[width=3.0in]{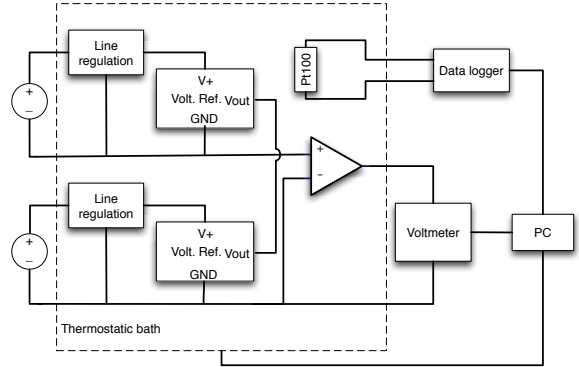}
\caption{Scheme of the experimental setup.}
\label{Fig:Setup}
\end{figure}

Ideally, measuring the noise and variability of the output voltage would require a ultra-stable reference (with a better voltage stability), so that a direct comparison could be done with the device under test (DUT). Since we do not have such a reference, our experimental setup is based on the comparison of two identical, yet independent, voltage references. The noise power of the differential measurement is then assumed to be twice the intrinsic noise power of each DUT (see figure~\ref{Fig:Setup}).

The voltage references have trimming capabilities to adjust the output voltage to the desired level. Nevertheless, the external trimming circuit induces an increase of the thermal drift (see e.g. [Fledderman 2009]) and was therefore not implemented. 
However, even in the absence of fine tuning, the voltage difference between the DUT is of the order of a few millivolts at most and the expected fluctuations are at the $\mu$V level. Therefore a low noise, proximity amplifier is required to increase the signal to noise ratio. The zero-drift, ultra-low noise ADA4528 operational amplifier, mounted as a non-inverting amplifier (with a gain of 401) was used for this purpose. 

For these very precise measurements, a direct comparison of output levels of the voltage references was not suitable, because of the finite common mode rejection of the measurement chain. The adopted solution was to connect the output pins of the 2 DUTs, while keeping the ground pin floating. The amplifier ground pin is then connected to one of the DUT's ground, while its non-inverting input is connected to the other DUT's ground pin. This configuration is effectively equivalent to measuring the output voltage of references with opposite nominal levels connected in series. 

The power supply is provided by a dual output Agilent E3649A (for the DUTs) and an Agilent E3631A (for the amplifier). 

Special care has been taken to select resistors and capacitance with very high thermal stability. E.g. by using high precision resistors, the dominant bias of he amplifier is thought to be the temperature dependance of the offset voltage, estimated to be equivalent to less than 5~$10^{-3}$~ppm/${}^\circ$C at the input of the amplifier.

The amplifier output is read by an AGILENT 34411A multimeter at a sampling rate of 50 Hz (i.e. averaging over 1 cycle of the power line frequency). The data are transferred to a PC running Labview and stored on the hard disk.

The temperature of the voltage references and the amplifier has to be strictly controlled and recorded to estimate its contribution and maintain its influence below the intrinsic noise of the reference. With a maximum thermal drift of 5~ppm/K, the thermal stability should be well below 0.1~K/$\sqrt{\text{Hz}}$ to reach the expected performance of eLISA voltage references (1~ppm/$\sqrt{\text{Hz}}$). For precise measurements, it is also important to keep the two voltage reference at the same temperature and minimize the thermal gradients between the temperature sensor and the DUTs. Consequently, the two electronics boards holding the voltage references and the amplifier were put into a bath of perfluoropolyether heat transfer fluid (Galden HT 110) which has a high resistance (about $10^{15}\ \Omega\cdot\text{cm}$) and a low viscosity (hence a good temperature homogeneity). This configuration also prevents any influence of the hygrometry and reduces the thermo-electric voltages at the soldered junctions on the electronics boards.

The fluid is thermally regulated using a thermostatic bath LAUDA Proline RP 845 C. The value of the in-loop thermistor (Pt100) is read and the set point remotely controlled from the acquisition computer. The value of this Pt100 is sampled at a rate of 2~Hz. For long term stability measurements, the temperature of the bath is set to 23${}^\circ$C. An additional sinusoidal modulation with amplitudes of 6 or 10 ${}^\circ$C peak-to-peak and a frequency of 0.2~mHz was also added for the measurement of the temperature coefficients (see below).

Another (out-of-loop) Pt100 sensor is placed on the electronic board, near to one of the DUT. Its value is monitored thanks to an AGILENT 34980A measurement unit (associated with a 34921A multiplexer) at a sampling rate of 5 Hz. The noise level of this temperature measurement is at the mK/$\sqrt{\text{Hz}}$ level down to 0.1 mHz. 

The thermostatic bath is placed into a Faraday cage, which is not - in this particular case - meant to protect the measurement from EM perturbations. Being thermally insulated, this cage allowed us to keep the voltmeter and the DUT at short distance, while thermally decoupling the voltmeter stability from the heat generated by the thermostatic bath.

The cage, as well as the different apparatus are placed in a air-conditioned room, with a thermal stability of about 1~K peak-to-peak around 22${}^\circ$C.

\section{Procedures and experimental results}

\subsection{Noise induced by the acquisition setup}

In order to estimate the noise level induced by the apparatus, two measurements have been performed : the estimation of the noise of the acquisition chain, and of the noise induced by the instability of the power supplies. 

\begin{figure}[!t]
\centering
\includegraphics[width=3.0in]{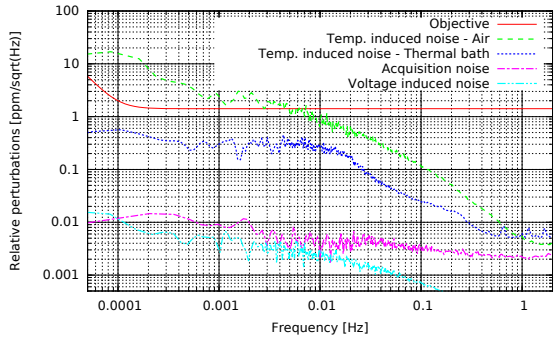}
\caption{Estimation of the measurements noise levels induced by the experimental setup.}
\label{Fig:MeasNoises}
\end{figure}

For the first measurement, the two inputs pins of the amplifier were connected together and the residual output voltage recorded by the voltmeter. The data are rescaled to ppm variations at the reference output by dividing by the gain of the amplifier (401) and by 5V, which is the nominal voltage level of all references, except for the AD587 (10~V). The thermal bath is regulated at 23${}^\circ$C and the duration of this measurement set to 20~000~s. After subtraction of the mean value (offset), the amplitude spectral density (ASD) of the recorded voltage is computed on a logarithmic frequency axes, using the procedure described in \cite{Trobs2006120}. The noise of the acquisition chain is between $10^{-3}$ and $2\cdot10^{-2}$~ppm/$\sqrt{\text{Hz}}$ from 0.1~mHz to 25~Hz, i.e. 2 orders of magnitude lower than the stability goal (see figure \ref{Fig:MeasNoises}). An (constant) offset equivalent to 0.15 ppm was observed during these measurements.

The stability of the power supplies have been measured by directly connecting the multimeter between the ground and the output pin of a voltage regulator (LM317), while keeping the rest of the system unchanged. In that particular case, measuring the difference between two regulators was not found to be adequate, since any common mode would then be suppressed (e.g. due to the same sensitivity to temperature), while it may have different effects on the reference voltage due to different line regulation coefficients. 
The obtained amplitude spectrum density is then multiplied by the highest line regulation coefficient from table \ref{Tab:ManChar} (20 ppm/V) and $\sqrt{2}$ to account for two independent source of noise on the differential measurement. This noise is represented on figure \ref{Fig:MeasNoises}. The measurement is however at the limit of the multimeter resolution and should be considered as an upper level of the noise due to voltage regulation.

The temperature-induced noise is estimated by multiplying the ASD of the bath temperature by twice the highest temperature sensitivity from table \ref{Tab:ManChar} (5 ppm/${}^\circ$C). The factor 2 assumes the worst case of voltage references with opposite temperature coefficients. The result on figure \ref{Fig:MeasNoises} shows that this noise level is about one order of magnitude below the eLISA requirements. For comparison, the same measurements has been done 'in air'~: without temperature regulation, the contribution of temperature fluctuations is likely to exceed the stability requirements. 
The actual temperature coefficient for each reference has nevertheless been measured (see below) and their contribution properly compared to the obtained results.

In the present experiment, the most important noise source seems therefore associated to temperature fluctuations, while the influence of the acquisition noise and line regulation are negligible.

\subsection{Temperature coefficients}

\begin{table}
\centering 
\renewcommand{\arraystretch}{1.0}
\caption{Temperature coefficients of the different tested voltage references. TC refers to the coefficients of one DUT, $\delta $TC to the differential measurement between the 2 DUTs, and TC${}_{m}$ the maximum value of the temperature coefficients as taken from the manufacturer's data sheet.} 
\begin{tabular}{c l r r r}
\hline
 Number & Model &   TC & $\delta$TC & TC${}_{m}$\\
  &  &   [ppm/K] & [ppm/K] &  [ppm/K] \\
\hline
1 & AD587UQ & 8.30 & -2.22 & 5\\
2 & ADR445BRZ & 1.9 & 1.72 &  3 \\
3 & ADR435BRZ & -3.1 & -1.8 & 3 \\
4 & MAX6126AASA50 & 0.47 & -0.45 & 3 \\
5 & LTC6655BHMS8-5 & -2.90 & 0.69 & 2\\
6 & MAX6350CSA+ & 0.32 & 0.12 & 1\\
7 & LT1021BCN8-5 & -0.18 & -0.67 & 5\\
8 & VRE305AD & 2.0 & -0.85 & 0.6\\
\hline
\end{tabular}
\label{Tab:ThermalCoef}
\end{table}

The temperature coefficients given by the manufacturers are usually computed using the box method. In this method, the maximum amplitude (peak-to-peak) variation of the output voltage is measured over the full operating temperature range of the device. The temperature coefficient is then the ratio of the voltage range divided by the temperature range, expressed in ppm of the nominal output voltage. Since the voltage references are usually temperature corrected up to the second order dependance, the evolution of the output voltage as a function of temperature of the reference is often 'S-shaped', whose slope at the operating temperature can be significantly different from the value given by the box method.

In the present study, the actual thermal coefficient of the voltage references and its influence on the stability measurement have been studied, in order to compare the temperature-induced fluctuations with stability of the output voltage.

The output voltage level of one of the two reference put in the thermostatic bath was, first, directly recorded with the multimeter (thus by-passing the amplifier), while the temperature of the bath is cycled (sinusoid) from 18 to 28 ${}^\circ$C, with a period of 5~000~s. The acquisition time was set to 20~h for these measurements. The recorded voltage output and the actual temperature of the bath were both fitted using a sinusoid model added to a second order polynomial. The polynomial was used to take into account any long term drift of the measurements. The ratio of the fitted voltage amplitude to the fitted temperature amplitude gives the temperature coefficient of the DUT, which is then rescaled to ppm, using the nominal level of the voltage reference.

\begin{figure}[!t]
\centering
\includegraphics[width=3.0in]{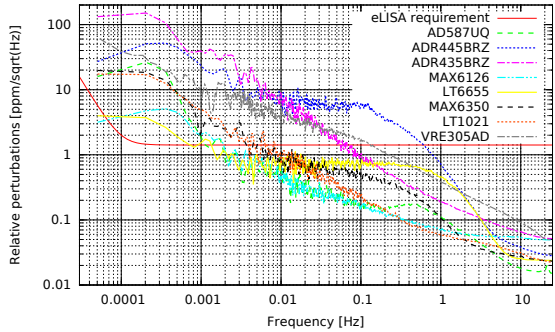}
\caption{Noise spectra computed from 20 ks acquisition.}
\label{Fig:ShortMeas}
\end{figure}

\begin{figure*}[t]
\centering
\subfloat[AD587UQ]{\includegraphics[width=3in]{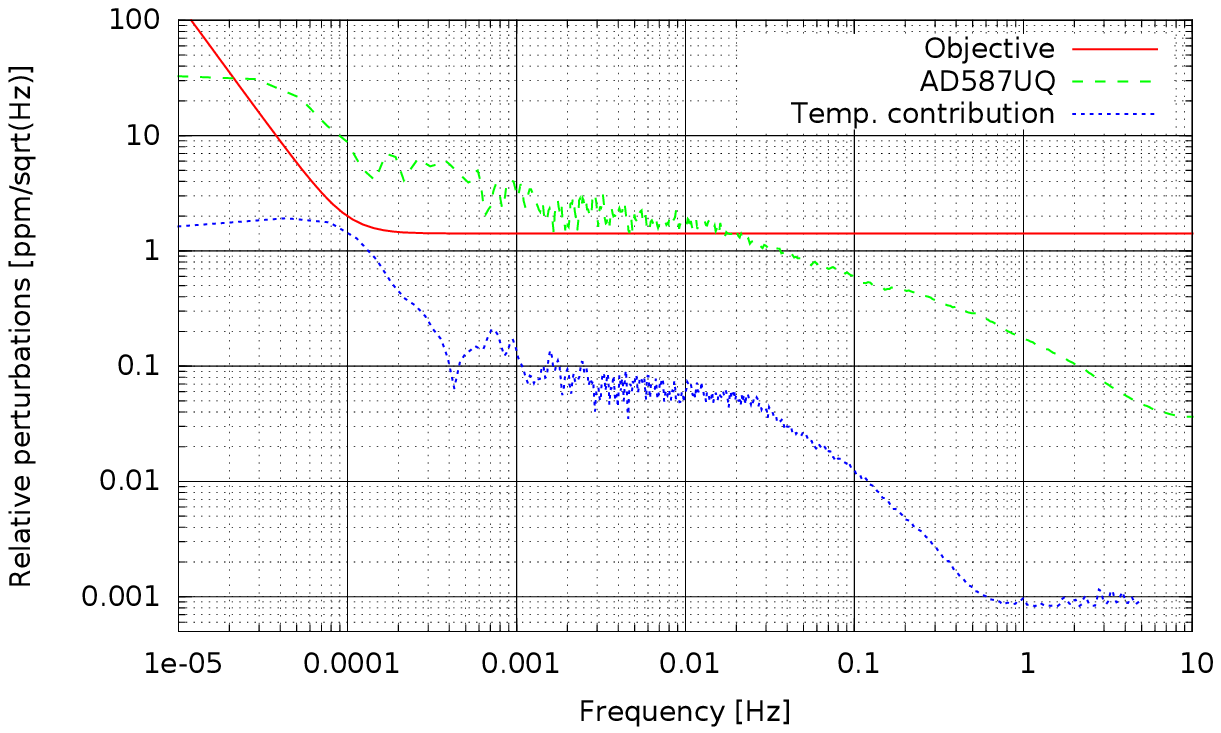}
\label{Fig:Ref1Noise}}
\hfil
\subfloat[MAX6126]{\includegraphics[width=3in]{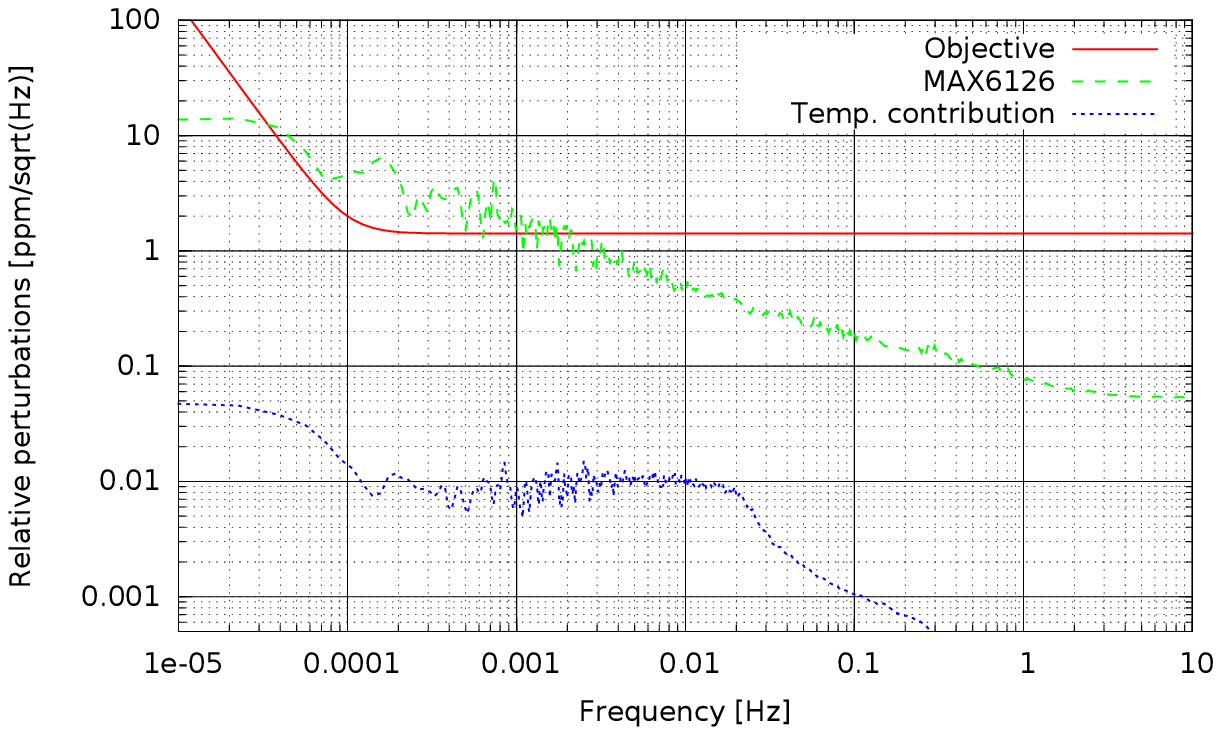}
\label{Fig:Ref4Noise}}
\hfil
\subfloat[LTC6655]{\includegraphics[width=3in]{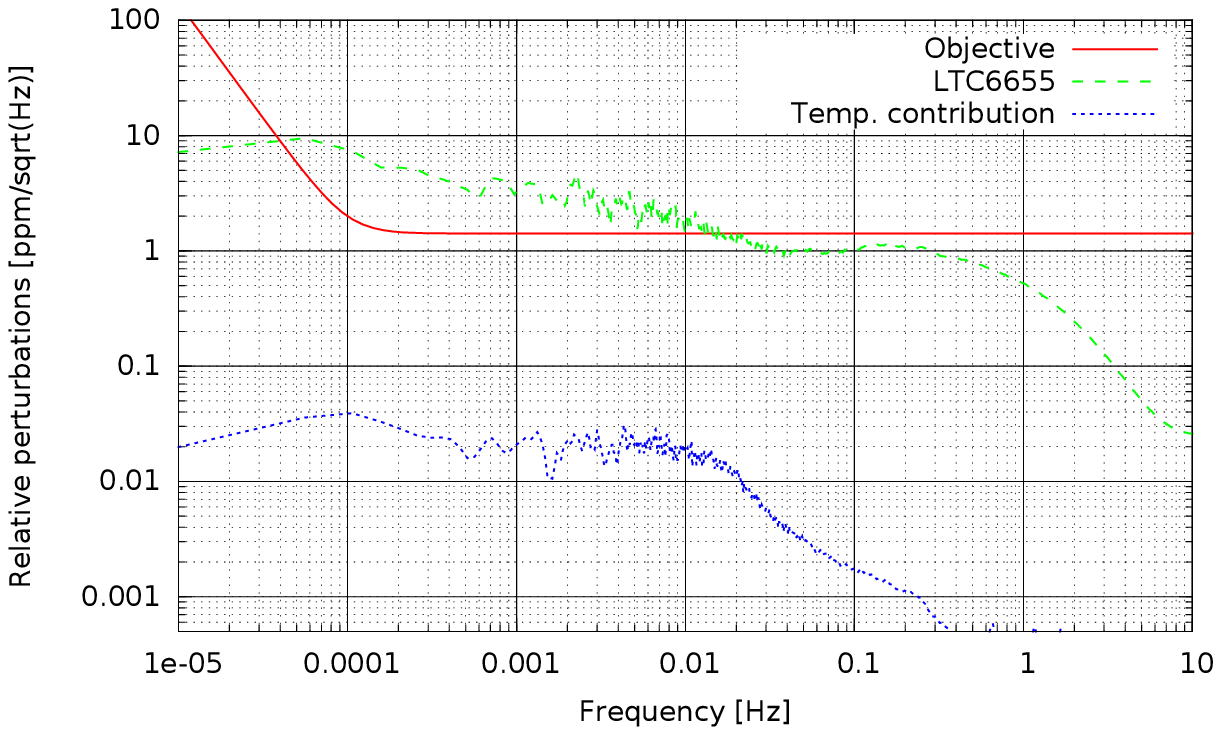}
\label{Fig:Ref5Noise}}
\hfil
\subfloat[MAX6350]{\includegraphics[width=3in]{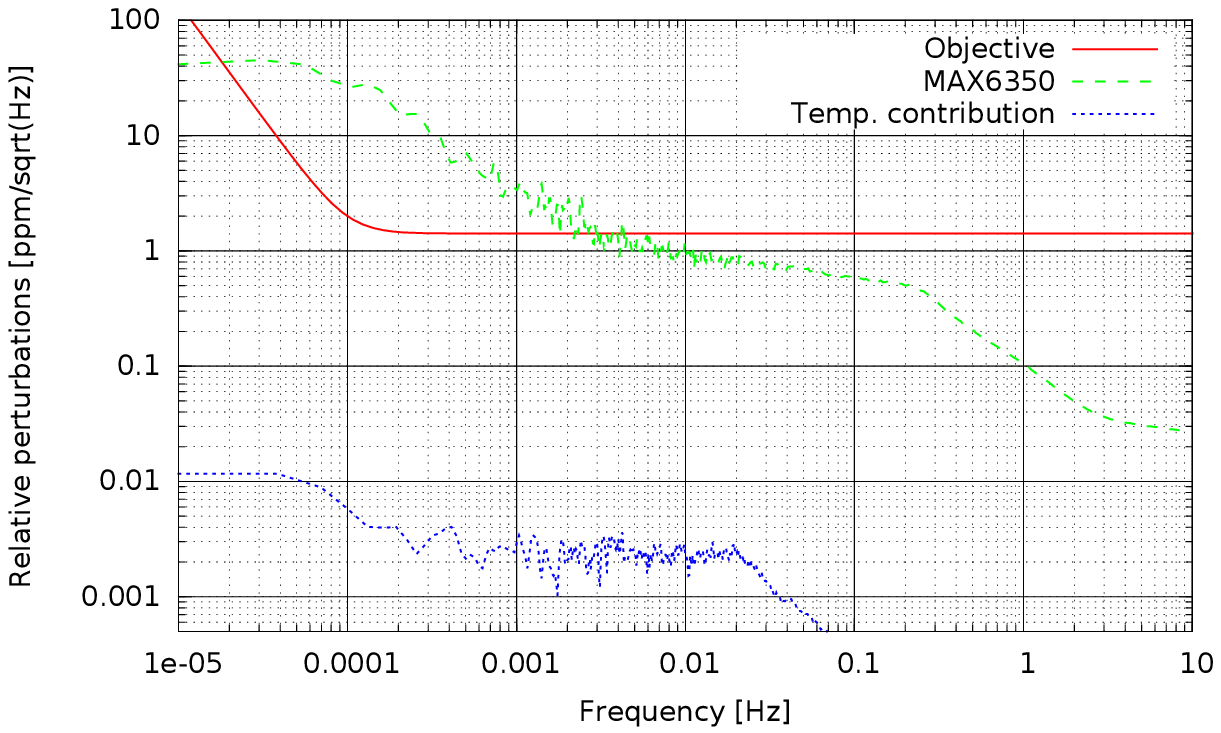}
\label{Fig:Ref6Noise}}
\hfil
\subfloat[LT1021]{\includegraphics[width=3in]{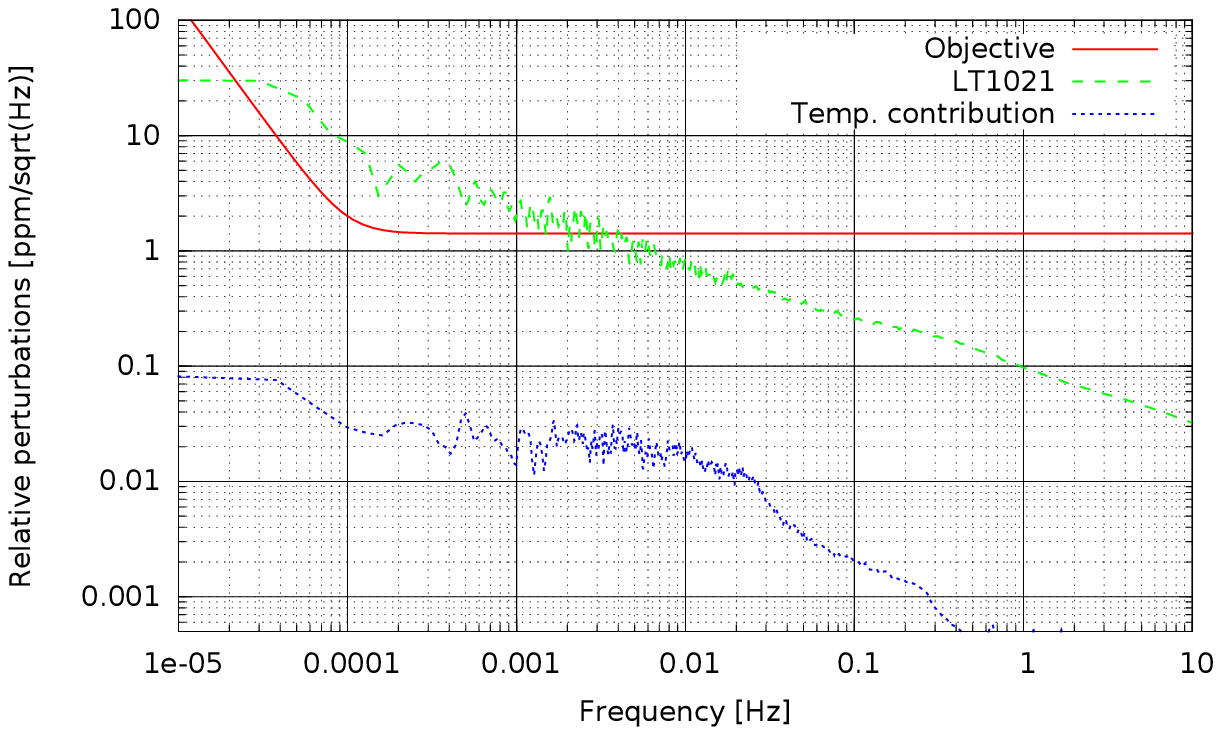}
\label{Fig:Ref7Noise}}
\caption{Noise spectra and temperature contributions for selected voltage references.}
\label{Fig:LongTermSpectra}
\end{figure*}

In a second step, the difference of the two voltage reference was recorded (after being amplified), while the thermal bath is cycled (6${}^\circ$C amplitude peak-to-peak, period of 5~000~s). The same analysis is then performed, but taking into account the amplifier gain. This measurement gives therefore the differential sensitivity of the voltage reference to temperature fluctuations. The signal to noise ratio of both measurements was sufficiently high to have error bars around 5\% typically.  The temperature coefficient of the second voltage reference can be deduced from these 2 measurements by subtracting the differential measurement to the temperature coefficient estimated with the direct measurement.
The results of this analysis are summarized in table \ref{Tab:ThermalCoef}.

The obtained values are globally compatible with the data sheet characteristics. The few discrepancies could be explained by the different (box and sine) analysis methods. However, with only two samples of each model, the purpose of this study is not to infer any general assessment on the voltage reference characteristics but to compute the influence of temperature on the intrinsic stability measurements.

\subsection{Long-term stability}

The long term stability is estimated by recording simultaneously the amplified differential voltage level and the temperature of the bath. The bath is kept at a constant temperature of 23${}^\circ$C during the acquisition. The contribution of the temperature to the voltage instability is estimated by multiplying the temperature amplitude spectral density by the differential temperature coefficient (see previous section).

Measurements have first been done on 'short' acquisition time (about 20 ks) in order to select the most promising references.
The results of this first screening is represented on figure \ref{Fig:ShortMeas}

From this figure, two voltage references (ADR445BRZ and ADR435BRZ, the only two tested devices based on the XFET technology) are significantly above the eLISA requirements, with a steep slope. The VRE305AD has roughly a 1/$\sqrt{\text{f}}$ evolution, but crosses at the relatively high frequency of 0.1~Hz.

The other references have equivalent noise performance around 1~mHz and appear better or close to the eLISA requirements down to this frequency. In order to get more significant results, these voltage references have been recorded on longer timescales, namely about 55~hrs.
The different noise spectra are represented on figure~\ref{Fig:LongTermSpectra}.

The AD587UQ reference is marginally compatible with the requirements down to a few mHz and is about 4 times the limit at 0.5 mHz. Having the largest differential temperature coefficient (see table \ref{Tab:ThermalCoef}), the contribution of temperature fluctuations to the output voltage is close to the eLISA requirements around 0.1 mHz.

The MAX6126 device appears compatible with the requirements on the whole frequency range, except between 0.05 mHz and 1 mHz. The influence of the temperature is negligible for this device. It appears to be the voltage reference with the lowest noise in the frequency range of eLISA among the samples tested in this work.

Similarly, the temperature fluctuations have negligible effect on the output stability of the LTC6655. The noise performance is comparable to the AD587 (slightly better below 0.1 mHz).  With this noise performance, comparable to the best buried zener tested, the LTC6655 voltage reference is an interesting alternative using another technology (bandgap).

As for the LT1021, its noise curve crosses the eLISA requirement at 0.5 mHz and is about 5 times the goal at 0.1 mHz. Its performance is comparable to the AD587UQ, but has a much lower temperature dependance.

All of these references, except  MAX6350, exhibit a typical 1/f behavior on the power spectral density (1/$\sqrt{\text{f}}$ in amplitude)  below 10~mHz which seems to indicate an intrinsic electronics noise. The MAX6350 device has a stepper slope (about 1/f in amplitude), which could be due to a slow drift of the voltage level.

\section{Conclusion}

The output stability of height - off the shelf - voltage references have been tested  in the frequency range from $10^{-5}$ to 10 Hz. None of the tested voltage reference meet the eLISA requirement, although the MAX 6126 is almost compatible, with a very low temperature dependance. The AD587UQ and LT1021 devices are buried zener references that could be interesting for eLISA. These results confirm earlier studies performed down to $10^{-4}$~Hz\cite{Fleddermann:2009p9044}. Another promising candidate, using bandgap technology is the LTC6655 voltage reference. On the other hand, the tested XFET technologies exhibit a relatively large voltage noise below 0.1~Hz.


%



\section*{Acknowledgment}
The authors wish to thank M. Filippo Marliani from the European Space Research and Technology Centre (ESTEC) at the European Space Agency (ESA) for useful discussion and advices in setting up this experiment.

\ifCLASSOPTIONcaptionsoff
  \newpage
\fi



\bibliographystyle{IEEEtran}
\bibliography{Biblio}
%



%

\newpage

\begin{IEEEbiography}{Hubert Halloin}
received the PhD degree in instrumentation for astrophysics from the University Paul Sabatier, Toulouse, France, in 2003.
 
Since 2005, he has been with the AstroParticle and Cosmology laboratory at the University paris Diderot, Paris, France, where he is working as an assistant professor. His research fields are with the development of interferometric instrumentation for space-based detectors, especially the eLISA project.

\end{IEEEbiography}

\begin{IEEEbiography}{Pierre Prat}
received a Diploma in Electrical Engineering from the INSA (National Institute of applied sciences) of Rennes, France in 1986.
In 2004, he joined the AstroParticle and Cosmology laboratory at the University Paris Diderot, Paris, France, where he is working as a research engineer in Electronics.
His current research interests are in the fields of low noise electronic system design for space-based detectors, especially the eLISA project.
\end{IEEEbiography}

\begin{IEEEbiography}{Julien Brossard}
 received the PhD degree in instrumentation for astrophysics from the University Paul Sabatier, Toulouse, France, in 2002. Since 2005, he is working at CNRS as a Research Engineer. From 2005 to 2012 he has been working in Particle Accelerator Physics at the Linear Accelerator laboratory (CNRS, Paris-Sud XI Univerisity, Orsay, France). Since 2012, he joined the AstroParticle and Cosmology laboratory (CNRS/CEA/Paris Diderot University/Paris Observatory, Paris, France). Currently, he is the project manager for the eLISA (detection and characterization of gravitational waves) hardware activities at APC and the system engineer for the QUBIC (detection and characterization of CMB B-mode polarization) project.
\end{IEEEbiography}







\end{document}